\documentclass{emulateapj}
\usepackage{graphics}

\slugcomment{Accepted by ApJ}

\begin{document}


\shortauthors{Temim et al.}

\shorttitle{``Kes 75 with \textit{Spitzer}, \textit{Chandra}''}

\title{INFRARED AND X-RAY SPECTROSCOPY OF THE Kes 75 SUPERNOVA REMNANT SHELL: CHARACTERIZING THE DUST AND GAS PROPERTIES}

\author{TEA TEMIM\altaffilmark{1,2}, PATRICK SLANE\altaffilmark{3}, RICHARD G. ARENDT\altaffilmark{1,4}, AND ELI DWEK\altaffilmark{1}}

\altaffiltext{1}{Observational Cosmology Lab, Code 665, NASA Goddard Space Flight Center, Greenbelt, MD 20771, USA}
\altaffiltext{2}{Oak Ridge Associated Universities (ORAU), Oak Ridge, TN  37831, USA; tea.temim@nasa.gov}
\altaffiltext{3}{Harvard-Smithsonian Center for Astrophysics, 60 Garden Street, Cambridge, MA 02138, USA}
\altaffiltext{4}{CRESST, University of Maryland-Baltimore County, Baltimore, MD 21250, USA}

\begin{abstract}

We present deep \textit{Chandra} observations and \textit{Spitzer Space Telescope} infrared (IR) spectroscopy of the shell in the composite supernova remnant (SNR) Kes 75 (G29.7-0.3). The remnant is composed of a central pulsar wind nebula and a bright partial shell in the south that is visible at radio, IR, and X-ray wavelengths. The X-ray emission can be modeled by either a single thermal component with a temperature of $\sim$ 1.5 keV, or with two thermal components with temperatures of 1.5 and 0.2 keV. Previous studies suggest that the hot component may originate from reverse-shocked SN ejecta. However, our new analysis shows no definitive evidence for enhanced abundances of Si, S, Ar, Mg, and Fe, as expected from supernova (SN) ejecta, or for the IR spectral signatures characteristic of confirmed SN condensed dust, thus favoring a circumstellar or interstellar origin for the X-ray and IR emission. The X-ray and IR emission in the shell are spatially correlated, suggesting that the dust particles are collisionally heated by the X-ray emitting gas. The IR spectrum of the shell is dominated by continuum emission from dust with little, or no line emission. Modeling the IR spectrum shows that the dust is heated to a temperature of $\sim$ 140~K by a relatively dense, hot plasma, that also gives rise to the hot X-ray emission component. The density inferred from the IR emission is significantly higher than the density inferred from the X-ray models, suggesting a low filling factor for this X-ray emitting gas. The total mass of the warm dust component is at least $1.3\times10^{-2}\:M_{\odot}$, assuming no significant dust destruction has occurred in the shell.
The IR data also reveal the presence of an additional plasma component with a cooler temperature, consistent with the 0.2 keV gas component. Our IR analysis therefore provides an independent verification of the cooler component of the X-ray emission. The complementary analyses of the X-ray and IR emission provide quantitative estimates of density and filling factors of the clumpy medium swept up by the SNR.

\end{abstract}


\section{INTRODUCTION} \label{intro}

The efficiency of dust formation in supernovae (SNe) and its subsequent evolution are still not well understood. Large discrepancies exist between theoretical predictions for dust formation in SNe and dust masses measured from recent infrared (IR) observations of supernova remnants (SNRs). Theoretical models based on the classical nucleation theory \citep{noz03,koz09} and the chemical kinematic approach for the formation of molecular precursors to dust in SN ejecta \citep{che11}
predict that the total dust mass produced per SN explosion should be 0.1--0.7 $M_{\odot}$.  A large fraction of this dust may be destroyed by erosion and sputtering between the forward and reverse shocks, but the surviving mass is still expected to be 0.04--0.2 $M_{\odot}$ \citep[e.g.,][]{dwe08,koz09}. In addition, dust evolution models show that an average SN would need to produce 0.1--1.0 $M_{\odot}$ of dust in order to explain the large quantities of dust observed in high-redshift galaxies \citep[e.g.,][]{mic10}. However, even such large production efficiencies may not be sufficient to explain the mass of dust in these objects \citep{dwe11,dwek11,val11,gal11a, gal11b}. 

In recent years, observations of SNRs with \textit{Spitzer} have provided mass estimates of newly formed warm (70--200 K) dust for several remnants, but these estimates lie in the 0.02--0.1 $M_{\odot}$ range \citep[e.g.,][]{sug06,rho08,rho09,tem10}, significantly less than theoretical predictions. 
Recent \textit{Herschel} observations revealed 0.4--0.7 $M_{\odot}$ of cool dust in SN 1987A, the highest mass estimate for freshly formed dust up to date \citep{mat11}.
\textit{Spitzer}, \textit{Herschel} and \textit{AKARI} observations of Cas A imply that 0.08 $M_{\odot}$ of dust is present in the remnant \citep{rho08,sib10,bar10}, consistent with model calculations for the evolution of dust formed in the ejecta of Type IIb SNe \citep{noz10}. \textit{Spitzer} observations of G54.1+0.3 showed evidence for up to 0.1 $M_{\odot}$ of fresh dust surrounding the pulsar wind nebula \citep[PWN;][]{tem10}. However, a large fraction of this dust may be destroyed by the SNR reverse shock, so the total mass estimate still falls short of what theories predict.

\begin{figure}
\epsscale{1.2} \plotone{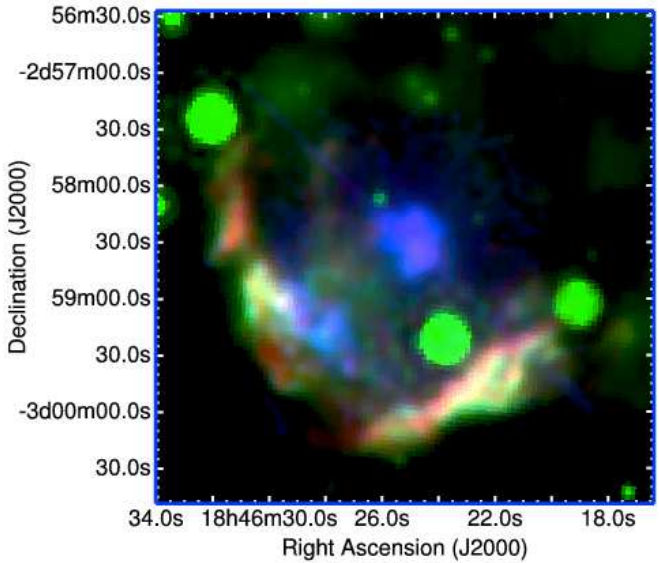} \caption{\label{3color}Three-color image of Kes 75, where the \textit{Chandra} X-ray emission is shown in blue, MIPS 24 $\micron$ emission in green, and radio emission in red. The IR emission shows a partial shell in the SE and SW regions and coincides with the X-ray and radio emission. The saturated point sources seen in green are foreground stars.}
\end{figure}

In order to determine the net amount of dust injected by SNe into the interstellar medium (ISM), it is also important to understand the dust destruction rates in SNR shocks. Hydrodynamic simulations of dust destruction in SNR reverse shocks suggest that 20\%--100\% of dust is destroyed by the shock, with the fraction depending on the grain species, ambient gas density, and the type of SN \citep{noz07,sil10}. The simulations show that grains smaller than 0.1 $\micron$ are either sputtered to smaller radii or completely destroyed. Recent modeling of the IR emission from Puppis A and the Cygnus Loop shows clear evidence of dust grain destruction in an SNR shock \citep{are10,san10}. The study of Large Magellanic Cloud (LMC) SNRs for both Type Ia and core-collapse SNRs by \citet{bor06} and \citet{wil06} shows that 40\% of the total mass in dust grains has been destroyed, including as much as 90\% of all grains smaller than 0.04 $\micron$. The inferred dust-to-gas mass ratios for these SNRs were several times smaller than the typically assumed LMC value of 0.025. In order to explain these results, either a much higher dust destruction rate or lower preshock dust-to-gas mass ratios is required \citep{bor06}. 

Kes 75 is an important subject for multi-wavelength studies of the dust and gas content of SNRs. Because of its relatively young age, it allows the possible distinction between ejecta and swept-up material giving rise to the X-ray and IR emission. Furthermore, it enables studies of the shocked dust and gas properties and grain destruction efficiencies in shocks. In a recent study of \textit{Chandra} and \textit{Spitzer} observations of Kes 75, \citet{mor07} suggest that the X-ray emission in Kes 75 arises from reverse-shocked ejecta, and also estimated the dust-to-gas mass ratio for the SNR shell to be $\sim 10^{-3}$, several times lower than what is typical for the Galaxy. However, these estimates were based only on the Multiband Imaging Photometer (MIPS) 24 $\micron$ measurement. Kes 75 was not detected at Infrared Array Camera (IRAC) wavelengths (3--8 $\micron$), suggesting that the small grains in the shell have been destroyed by shocks \citep{mor07}. Our new \textit{Spitzer} spectroscopy allows us to place more detailed constraints on the dust properties and mass by fitting the spectra from the shell directly.

Kes 75 is a Galactic SNR consisting of a PWN and a partial thermal shell. The PWN is powered by a 0.3 s pulsar (PSR J1846-0258) that has a  characteristic age of 723 yr, one of the youngest known \citep{got00}. An upper limit on its true age of 884 yr was derived from the measurement of the braking index, $n=2.65\pm0.01$ \citep{liv06}. The spin-down rate $\dot P=7.1\times10^{-12}\rm \: s \:s^{-1}$ leads to a spin-down luminosity of $\dot E=8.1\times10^{36}\rm \: erg \: s^{-1}$ and a strong surface magnetic field $B=5\times10^{13}\rm \: G$ \citep{got00}. The distance to Kes 75 is poorly constrained, with the estimates ranging from 5.1 to 21 kpc \citep{cas75,mil79,bec84,leah08}. Recent millimeter observations of CO line emission toward Kes 75 provide evidence for an association with an adjacent molecular cloud at a distance of $\sim 10.6 \rm \: kpc$ \citep{su09}. We adopt this distance throughout the paper, at which the radius of the SNR shell is equal to 5.7 $\rm pc$. The dependence of the derived gas and dust properties on the assumed distance will be discussed later in the paper.

Radio observations of Kes 75 show a partial SNR shell, $\sim$ 1$\farcm$5 in radius, with a spectral index $\alpha = 0.7$, and a flatter PWN component with $\alpha = 0.25$ \citep{bec76,hel03}. Here, $\alpha$ is defined such that $L_{\nu}\propto\nu^{-\alpha}$, where $L_{\nu}$ is the synchrotron luminosity. X-ray observations with \textit{ASCA} and \textit{Chandra} also showed the composite nature of the remnant with a morphology closely resembling the radio \citep{bla96,hel03}. \textit{Chandra} spectroscopy of the PWN shows that the emission softens with distance from the pulsar, with the photon index, $\Gamma$ ranging from 1.1 to 1.9 \citep{mor07, ng08}, where $\Gamma$ relates the number of photons $N$ and energy $E$ such that $dN/dE\propto E^{-\Gamma}$. The shell emission in Kes 75 is primarily concentrated in two regions in the southeast (SE) and the southwest (SW). \textit{Chandra} observations show that the spectra of the shell can be characterized by a two-temperature thermal model, possibly associated with the forward-shocked material and the reverse-shocked ejecta \citep{mor07}.

The first IR detection of Kes 75 with \textit{Spitzer} revealed a partial IR shell coincident with the shell at radio and X-ray wavelengths \citep{mor07}. \citet{mor07} used \textit{Chandra} and MIPS 24 $\micron$ observations, along with IRAC upper limits, to constrain the dust emission models and obtain a dust-to-gas mass ratio $M_{\rm dust}/M_{\rm gas} \sim 5\times10^{-4}$, more than an order of magnitude lower than that of the Galaxy. Since Kes 75 was only detected in the MIPS 24 $\micron$ band, the data were not sufficient to characterize the dust composition and place tight constraints on the dust temperature and mass. 

Here, we present \textit{Chandra} X-ray observations and \textit{Spitzer} IR spectroscopy of the shell in Kes 75 that allowed us to better characterize the observed dust and gas properties. Fitting of the X-ray spectra allows us to determine the properties of the thermal emission from the shell, while the modeling of the IR spectra from the shell provides the dust mass and temperature, and an independent estimate of the temperature and density of the heating plasma. The comparison of the X-ray and IR modeling results provides additional insight into the nature of the shell emission in Kes 75.

\section{OBSERVATIONS AND DATA REDUCTION} \label{obsv}

\subsection{\textit{Chandra}}

The \textit{Chandra} observations of Kes 75 were carried out with the Advanced CCD Imaging Spectrometer on 2006 June 5, 7, 9, and 12, under observation ID numbers 7337, 6686, 7338, and 7339, and an exposure time of 18, 55, 40, and 42 ks, respectively. The dataset was previously analyzed by \citet{mor07} and \citet{su09}, but here, we reanalyze the spectra using a different approach to background selection and fitting. The data were reprocessed and cleaned using CIAO v. 4.3, resulting in total clean exposure time of 152 ks.

\begin{figure}
\epsscale{1.1} \plotone{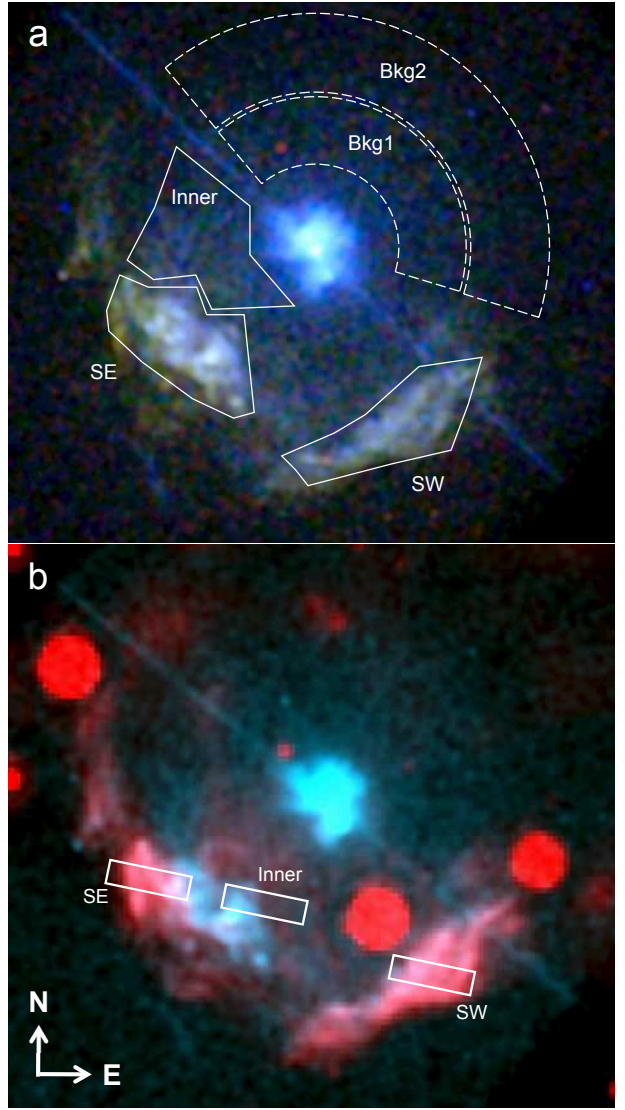} \caption{\label{3color2}(a) Three-color \textit{Chandra} X-ray image of Kes 75, with 0.3--2.0 keV emission in red 2.0--4.0 keV emission in green and 4.0--9.0 keV emission in blue. The extraction regions for X-ray source and background spectra are shown in white. (b) Two-color images of Kes 75 with MIPS 24 $\micron$ emission in red and \textit{Chandra} X-ray emission from 0.5--9.0 keV in cyan. The sub-slits of the IRS LL module used for the low-resolution IR spectral extraction are shown in white. We note that the regions labeled ``Inner'' in top and bottom panels do not spatially coincide.}
\end{figure}

Since the observations were carried out close in time and have similar chip orientation, we merged the individual data sets into a single event file. The merged image is shown in Figures \ref{3color} and \ref{3color2}. We chose extraction regions for the SE and SW rims similar to those in \citet{mor07} and \citet{su09}, with an additional region covering the inner SNR between the PWN and the southeastern rim of the shell. The extraction regions for the X-ray spectra are shown in Figure \ref{3color2}(a). 

\subsection{\textit{Spitzer} Spectroscopy}

\begin{deluxetable}{lcc}
\tablecolumns{3} \tablewidth{0pc} \tablecaption{\label{irfluxtab}MIPS 24 MICRON Fluxes}
\tablehead{
\colhead{Region} & \colhead{Flux (Jy)} & \colhead{Ext. Corr. Flux (Jy)} 
}
\startdata
SE Rim & 1.9 $\pm$ 0.4 & 5.1 $\pm$ 1.0 \\
SW Rim & 2.7 $\pm$ 0.5 & 7.0 $\pm$ 1.4 \\
\cutinhead{Emission from the Shell minus Inner SNR}
SE  Rim & 0.9 $\pm$ 0.2 & 2.2 $\pm$ 0.4 \\
SW Rim  & 1.3 $\pm$ 0.3 & 3.4 $\pm$ 0.7 \\
\enddata
\tablecomments{The uncertainties include MIPS calibration uncertainties, but do not account for the uncertainties in the extinction correction. The extinction correction was applied using the extinction curve of \citet{chi06} and a hydrogen column density of $N_{\rm H}=3.5\times10^{22} \: \rm cm^{-2}$.}
\end{deluxetable}

\begin{figure}
\epsscale{1.05} \plotone{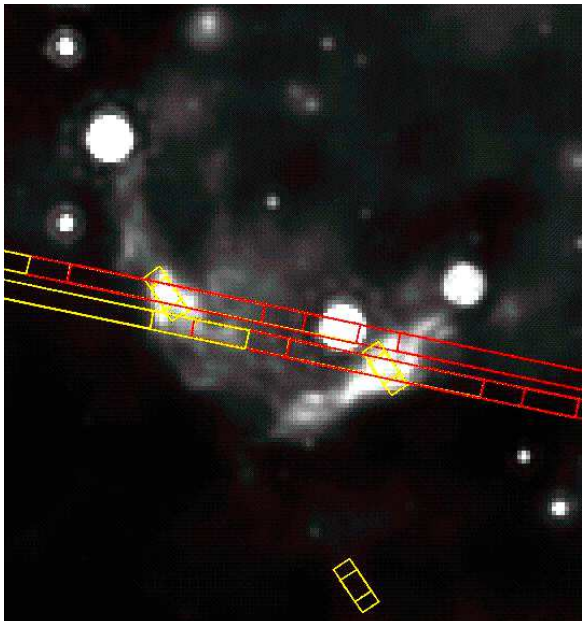} \caption{\label{slits}Positions of the IRS slits overlaid on the MIPS 24 $\micron$ image. The LL slits are the large slits oriented in the east/west direction and span the entire length of the SNR shell. The LH slits are the smaller slits shown in yellow, positioned at the east and west portions of the shell. The LH slits located south of the remnant were used for background subtraction.}
\end{figure}

\textit{Spitzer} IR spectroscopy of the shell in Kes 75 was carried out on 2008, June 5 (program ID 50447) using the long-low (LL) and long-high (LH) modules of the Infrared Spectrograph \citep[IRS;][]{hou04} that cover the range from 14--37 $\micron$. The locations of the IRS slits overlaid on the MIPS 24 $\micron$ image \citep{mor07} are shown in Figure \ref{slits}. The spectra were taken at positions $18\fh46\fm22\fs3$, -2$^{\circ}$59$\arcmin46\farcs9$ and 
$18\fh46\fm30\fs36$, -2$^{\circ}$58$\arcmin54\farcs9$, corresponding to the SW and SE portions of the shell, respectively. In order to measure the background emission for the LH module, we also placed a slit at position $18\fh46\fm24\fs03$, -3$^{\circ}$01$\arcmin49\farcs0$, south of the partial shell. All observations were taken with 10 cycles of 6 s exposures and the data were processed with the pipeline version S18.0.2.

The LH spectroscopy was cleaned with the IRS Rogue Pixel Mask Editing and Image Cleaning Software (IRSCLEAN1.9). The background spectrum was subtracted from the LH data and the Basic Calibrated Data of each nod position were combined. Spectra for the SW and SE portion of the shell were extracted using the Spitzer IRS Custom Extractor (SPICE) with full slit extractions and extended source calibration. For the LL modules, the spectra were extracted using the CUbe Builder for IRS Spectral Maps \citep[CUBISM;][]{smi07}, which allowed us to select sub-slits from the off-source nods for background subtraction. Since the background in the vicinity of Kes 75 is spatially variable (89--102 $\rm MJy \: sr^{-1}$ at 24 $\micron$), some off-source nod positions produced negative residuals in the background-subtracted spectra. The final background was produced by averaging spectra extracted from off-source sub-slits that did not result in negative residuals. The source spectra were extracted from sub-slits covering the shell emission and inner region on the SNR, and are shown in Figure \ref{3color2}(b). The rest of the analysis was carried out with the Spectroscopic Modeling, Analysis, and Reduction Tool \citep[SMART;][]{hig04}, including fitting of the emission lines.

\section{ANALYSIS AND RESULTS}\label{analysis}

\subsection{Multi-wavelength Morphology}\label{morph}

The three-color image of Kes 75, displayed in Figure \ref{3color}, shows the \textit{Chandra} X-ray emission in blue, MIPS 24 $\micron$ IR emission in green, and Very Large Array radio emission in red \citep{bec84}. The IR image shows a partial SNR shell, with the brightest emission concentrated in the SE and SW rims which coincides spatially with the X-ray and radio emission. The 24 $\micron$ image also shows fainter IR emission in the inner region of the SNR, between the PWN and the southern rim, that is more diffuse than the filamentary shell. Extinction-corrected 24 $\micron$ fluxes of the bright portions of the shell are measured to be 5.1 and 7.0 Jy in the SE and SW rims, respectively. We also measured the total flux in the shell using the diffuse IR emission inside of the SNR as background. Since the IR spectra suggest that the emission from the shell is composed of two distinct dust temperature components, the latter flux densities are estimates for the warm dust component that is concentrated in the shell (see Section \ref{irspec}).
The measured values are listed in Table \ref{irfluxtab}. As discussed in Section \ref{irspec}, the IR emission is dominated by dust continuum, with very little or no contribution from line emission. The radio synchrotron emission from the shell has a spectral index $\alpha$ = 0.7 and a flux density of $\sim$ 10 Jy at 1 GHz \citep{bec76}. When extrapolated to \textit{Spitzer} IR wavelengths, the contribution from the synchrotron emission is negligible.

The X-ray emission spatially correlates with the IR, as is best seen in Figure \ref{3color2}. The rims of the SNR shell are dominated by thermal emission with a range of temperatures. This is best seen in the X-ray three-color image in Figure \ref{3color2}(a). The colors show that some portions of the shell are dominated by softer X-ray emission, while other regions have a harder X-ray spectrum (blue). The SE portion of the shell appears particularly clumpy, with softer X-ray clumps embedded in more diffuse hard emission. Figure \ref{3color2} also shows that diffuse X-ray emission fills the inner SNR. \citet{su09} show that the spectrum of the diffuse emission north of the PWN is featureless and most likely non-thermal in nature. This emission likely represents the dust-scattered halo from the PWN. However, the emission from the inner SNR region south of the PWN is more enhanced than in the north, and shows evidence of thermal X-ray emission.


\subsection{X-Ray Spectroscopy}

The \textit{Chandra} X-ray spectra were extracted from regions shown in Figure \ref{3color2}(a). The best fit parameters are shown in Table \ref{xrayfitstab}. We also refitted the spectrum of the entire PWN, excluding the pulsar, with a power-law model and the Tuebingen--Boulder ISM absorption model XSTBABS. The spectrum contained $\sim$ 100,000 counts after subtracting a background outside of the SNR. The absorbed power-law model provided a good fit to the spectrum, giving a $\chi^2$ of 0.92. Using the XSTBABS absorption model, we find $N_{\rm H}=3.75\times10^{22}\: \rm cm^{-2}$, a slightly lower value than reported previously, and $\Gamma=1.98\pm0.3$, consistent with previous results. 

 \citet{su09} modeled the X-ray spectra from the SE and SW rims with a thermal component, plus a non-thermal one to account for possible dust scattering. Instead of using this approach, we assumed that the emission from the dust scattering halo is approximately symmetric about the PWN, and we chose background regions in the northern part of the shell that do not appear to be contaminated by thermal emission to subtract the scattered emission component. The background regions are shown in Figure \ref{3color2}(a). The region labeled ``Bkg1" was used as background for the ``Inner" region, while the region labeled ``Bkg2" was used as background for the SE and SW regions. The source spectra for the SE (SW) rim contained 37,000 (16,000) counts, while the total counts in the background, including emission from the dust scattering halo, was approximately 2,900 (2700) counts. The total number of source counts in the ``Inner" region was 8900, but the background in this case was significantly higher, approximately 5350 counts. The spectra from the SE and SW regions were grouped by 30 counts bin$^{-1}$, while the spectrum from the inner region was grouped by 100 counts bin$^{-1}$.
 
\renewcommand{\arraystretch}{1.5}
\begin{deluxetable*}{lcccc}
\tablecolumns{5} \tablewidth{0pc} \tablecaption{\label{xrayfitstab}\textit{Chandra}
Spectral Fitting Results} 
\tablehead{ \colhead{Parameter} &
\colhead{PWN} & \colhead{SE Shell} & \colhead{SW Shell}  & \colhead{Inner SNR}} 
\startdata
\cutinhead{VPSHOCK--One-component Model} \\
$N_H(10^{22}\rm cm^{-2})$ & $3.75^{+0.06}_{-0.06}$  & $2.93^{+0.02}_{-0.04}$ &  $3.12^{+0.03}_{-0.10}$  &  $3.58^{+0.22}_{-0.23}$ \\
Photon Index ($\Gamma$) & $1.98^{+0.03}_{-0.03}$ & \nodata  & \nodata  & \nodata \\
$kT (\rm keV)$ & \nodata & $1.48_{-0.02}^{+0.02}$ & $1.93_{-0.03}^{+0.23}$  & $2.47_{-0.23}^{+0.22}$   \\
$\rm \tau (s\:cm^{-3})$ & \nodata & $(1.46_{-0.11}^{+0.12})\times10^{11}$ &  $(7.36_{-0.72}^{+1.24})\times10^{10}$ & $(1.65_{-0.52}^{+1.11})\times10^{10}$  \\
$F \rm(unabs. \: flux\: (\rm erg\:cm^{-2}\:s^{-1}))$ & $3.6\times10^{-11}$ & $5.5\times10^{-11}$ & $3.2\times10^{-11}$ & 
$1.8\times10^{-11}$ \\
Normalization &  \nodata & $(7.7_{-0.4}^{+0.1})\times10^{-3}$ & $(2.81_{-0.26}^{+0.04})\times10^{-3}$ & $(8.1_{-2.3}^{+1.7})\times10^{-4}$ \\
Reduced $\chi^2$ statistic & 0.92 & 1.33 & 1.34 & 0.92 \\

\cutinhead{VPSHOCK--Two-component Model}
$N_H(10^{22}\rm cm^{-2})$ & \nodata & $3.33^{+0.18}_{-0.56}$ &  $4.00^{+0.14}_{-0.17}$  &  \nodata \\
$kT_1 (\rm keV)$ & \nodata & $1.46_{-0.04}^{+0.62}$ & $1.60_{-0.09}^{+0.09}$  & \nodata   \\
$\rm \tau_1 (s\:cm^{-3})$ & \nodata & $(1.09_{-0.10}^{+0.19})\times10^{11}$ &  $(5.23_{-0.86}^{+1.00})\times10^{10}$ & \nodata  \\
$F_1 \rm(unabs.\: flux \: (\rm erg\:cm^{-2}\:s^{-1}))$ & \nodata & $7.23\times10^{-11}$ & $6.1\times10^{-11}$ & \nodata \\
$\rm Normalization_1$ & \nodata & $(8.6_{-3.3}^{+0.3})\times10^{-3}$ & $(4.6_{-0.4}^{+0.4})\times10^{-3}$ & \nodata \\
$kT_2 (\rm keV)$ & \nodata & $0.20_{-0.03}^{+0.56}$ & $0.14_{-0.01}^{+0.02}$  & \nodata  \\
$\rm \tau_2 (s\:cm^{-3})$ & \nodata & $(2.97_{-2.42}^{+11.8})\times10^{9}$ & $(2.22_{-1.15}^{+1.35})\times10^{10}$ & \nodata \\
$F_2 \rm(unabs.\: flux \: (\rm erg\:cm^{-2}\:s^{-1}))$ &  \nodata & $3.4\times10^{-9}$ & $1.0\times10^{-8}$  & \nodata \\
$\rm Normalization_2$ & \nodata & $2.8_{-2.8}^{+2.5}$ & $14.3_{-6.3}^{+5.1}$ & \nodata \\
Reduced $\chi^2$ statistic & \nodata & 1.01 & 0.92 & \nodata \\

\cutinhead{VNEI Model}
$N_H(10^{22}\rm cm^{-2})$ & \nodata & $3.38^{+0.02}_{-0.01}$ &  $3.49^{+0.69}_{-0.03}$  &  $3.73^{+0.25}_{-0.24}$ \\
$kT_1 (\rm keV)$ & \nodata & $1.60_{-0.01}^{+0.02}$ & $2.08_{-0.37}^{+0.17}$  & $2.11_{-0.37}^{+0.55}$   \\
$\rm \tau_1 (s\:cm^{-3})$ & \nodata & $(5.70_{-0.33}^{+0.47})\times10^{10}$ &  $(3.14_{-0.27}^{+0.66})\times10^{10}$ & $(7.06_{-1.31}^{+1.93})\times10^{9}$  \\
$F_1 \rm(unabs. \: flux (\rm erg\:cm^{-2}\:s^{-1}))$ & \nodata & $4.0\times10^{-11}$ & $2.8\times10^{-11}$ & $2.6\times10^{-11}$ \\
$\rm Normalization_1$ & \nodata & $(6.80_{-0.07}^{+0.08})\times10^{-3}$ & $(2.5_{-0.2}^{+0.2})\times10^{-3}$ & $(1.0_{-0.3}^{+0.4})\times10^{-3}$ \\
$kT_2 (\rm keV)$ & \nodata & $0.24_{-0.01}^{+0.04}$ & $0.20_{-0.09}^{+0.09}$  & \nodata  \\
$\rm \tau_2 (s\:cm^{-3})$ & \nodata & $>1.5\times10^{13}$ & $>2.57\times10^{12}$ & \nodata \\
$F_2 \rm(unabs. \: flux \: (\rm erg\:cm^{-2}\:s^{-1}))$ &  \nodata & $2.1\times10^{-10}$ & $1.5\times10^{-10}$  & \nodata \\
$\rm Normalization_2$ & \nodata & $0.141_{-0.005}^{+0.007}$ & $0.11_{-0.5}^{+0.6}$ & \nodata \\
Reduced $\chi^2$ statistic & \nodata & 1.41 & 1.27 & 0.97 \\
\enddata
\tablecomments{The listed uncertainties are 1.6 $\sigma$ (90\% confidence) statistical uncertainties only. The apertures used in the spectral extractions are shown in Figure \ref{3color2} (top), and the spectra and fits are shown in Figures \ref{xrayspec1}--\ref{xrayspec3}. The fluxes were calculated in the 0.5--9.0 keV. The normalization of the thermal models is equal to $10^{-14}n_e n_{\rm H} V / 4\pi d^2 \: \rm cm^{-5}$, where $V$ is the volume of the emitting region and $d$ is the distance to the SNR.}
\end{deluxetable*}
\renewcommand{\arraystretch}{1.0}

 \begin{figure*}
\epsscale{1.0} \plottwo{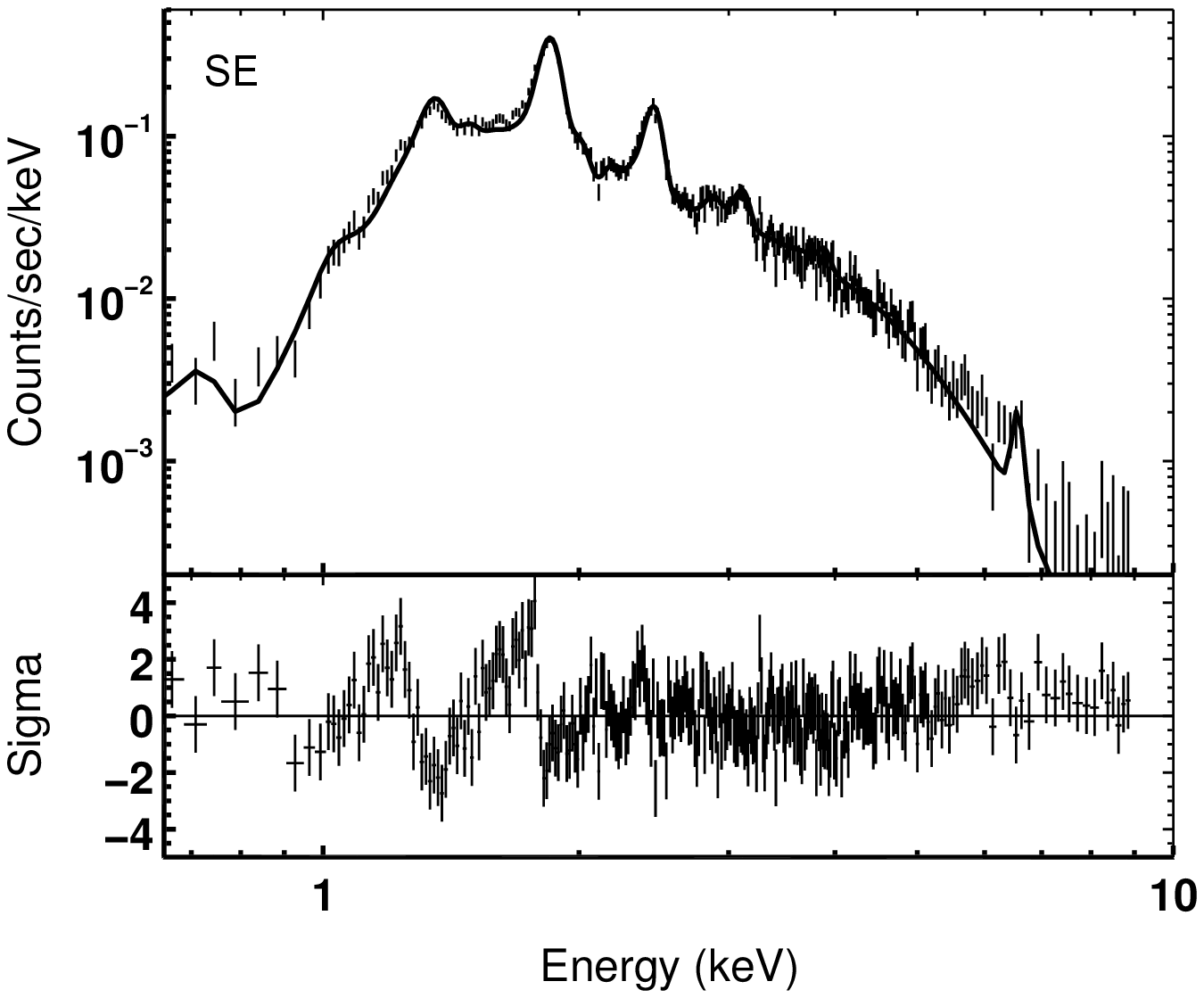}{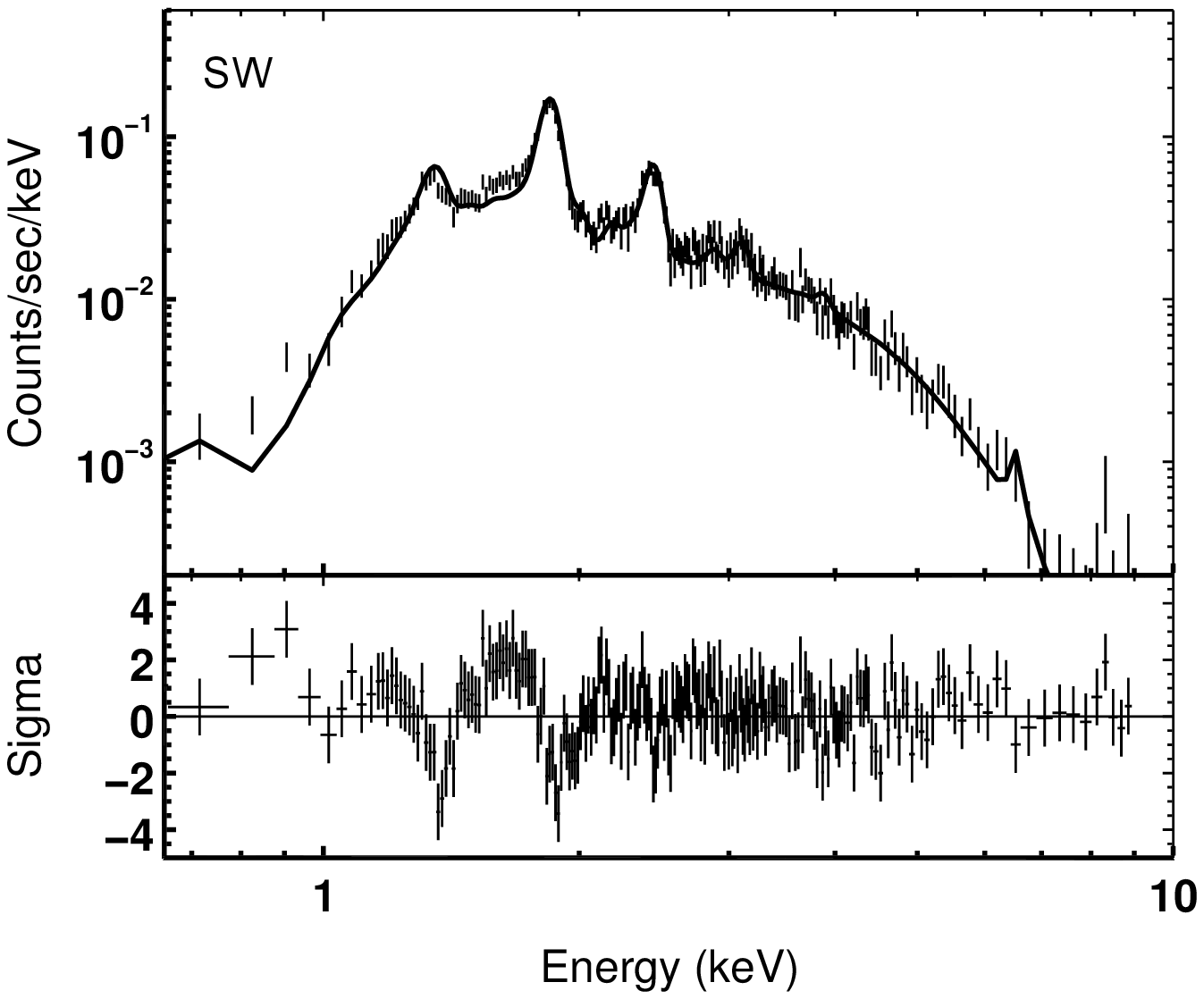} \caption{\label{xrayspec1} \textit{Chandra} X-ray spectra and best-fit models for the SE (left) and SW (right) regions of the shell. The extraction regions are shown in Figure \ref{3color2} and best-fit models listed in Table \ref{xrayfitstab}. The spectra were fitted with a single-temperature VPSHOCK model; 1.5 keV in the SE and 1.9 keV in the SW. Observed emission lines are from Si, S, Mg, Ar, and Fe.}
\end{figure*}

\begin{figure*}
\epsscale{1.0} \plottwo{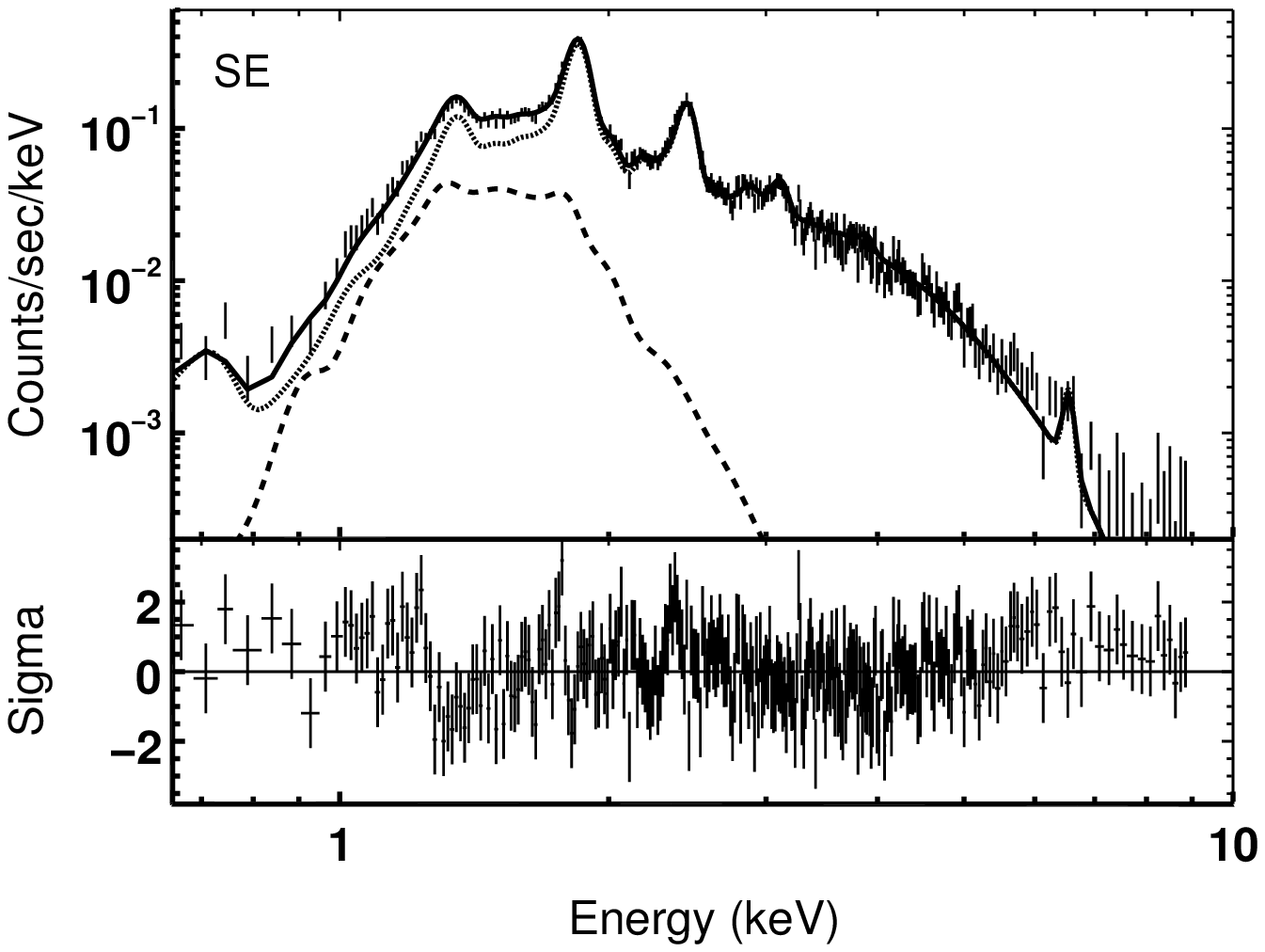}{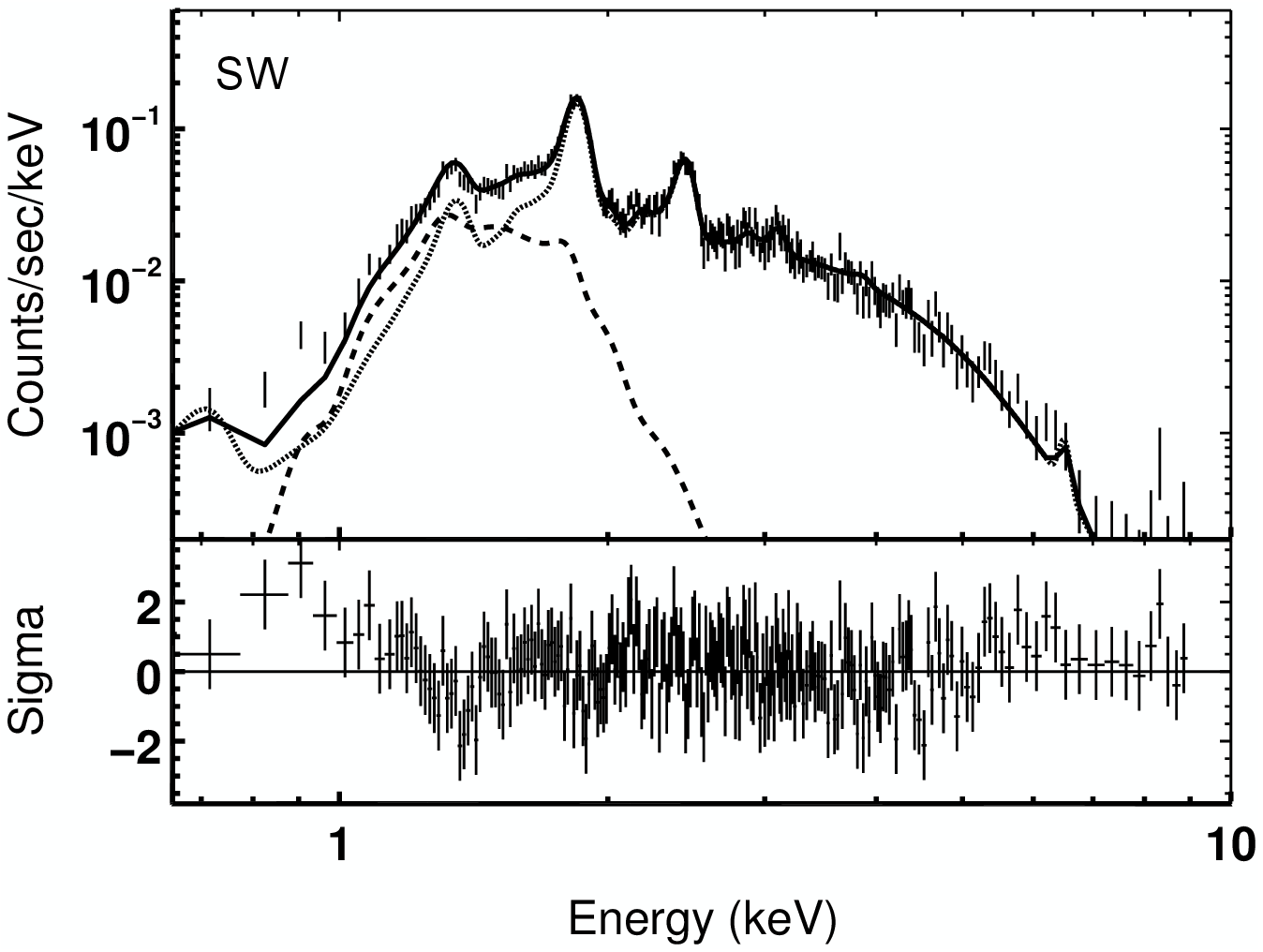} \caption{\label{xrayspec2} Same as Figure \ref{xrayspec1}, but with a two-temperature VPSHOCK model (solid line), composed of a hotter component with a temperature of 1.5 (1.6) keV (dotted line) and a cooler component with a temperature of 0.2 (0.14) keV (dashed line), for the SE (SW) shell.}
\end{figure*}

\begin{figure*}
\epsscale{1.0} \plottwo{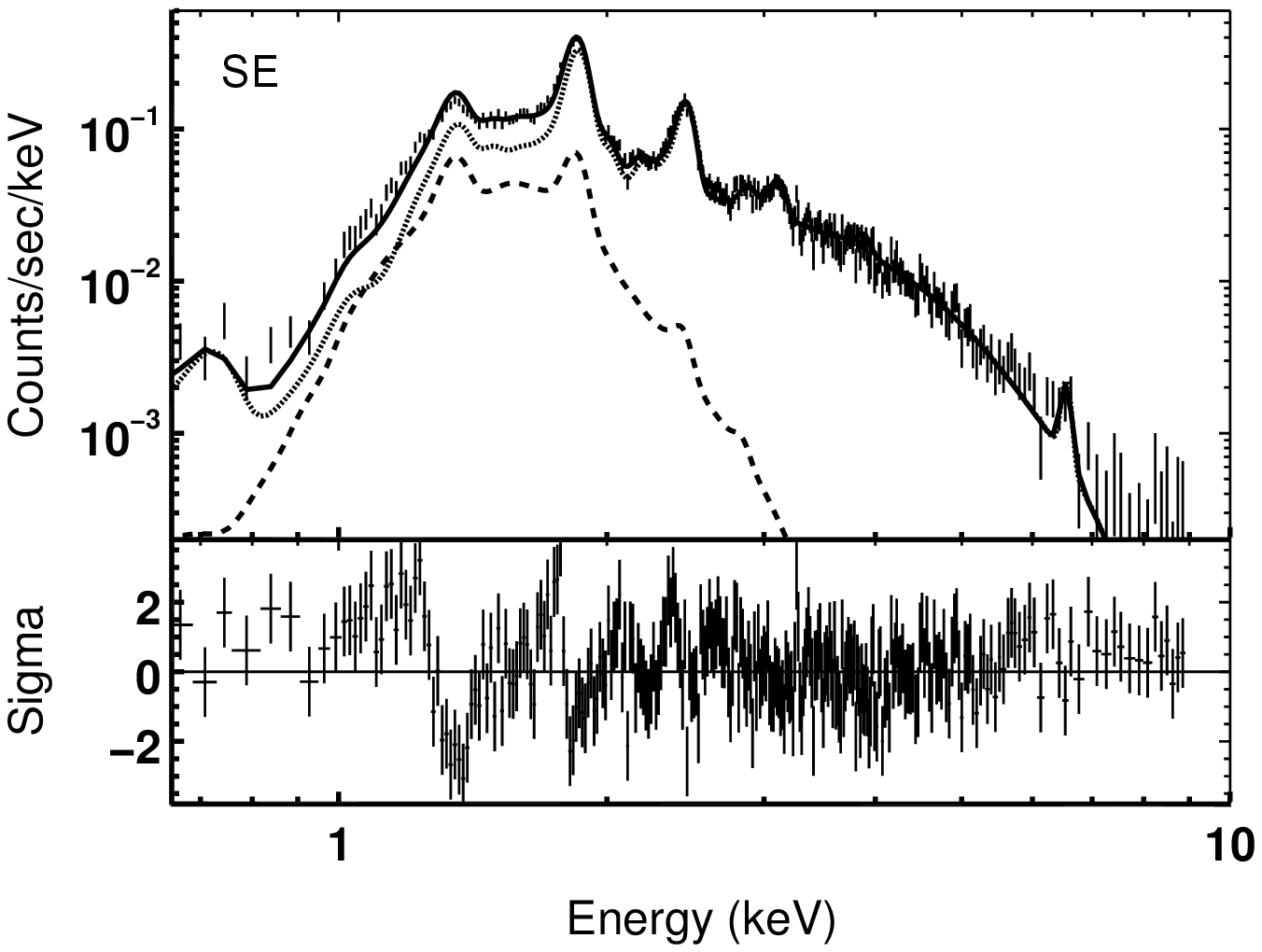}{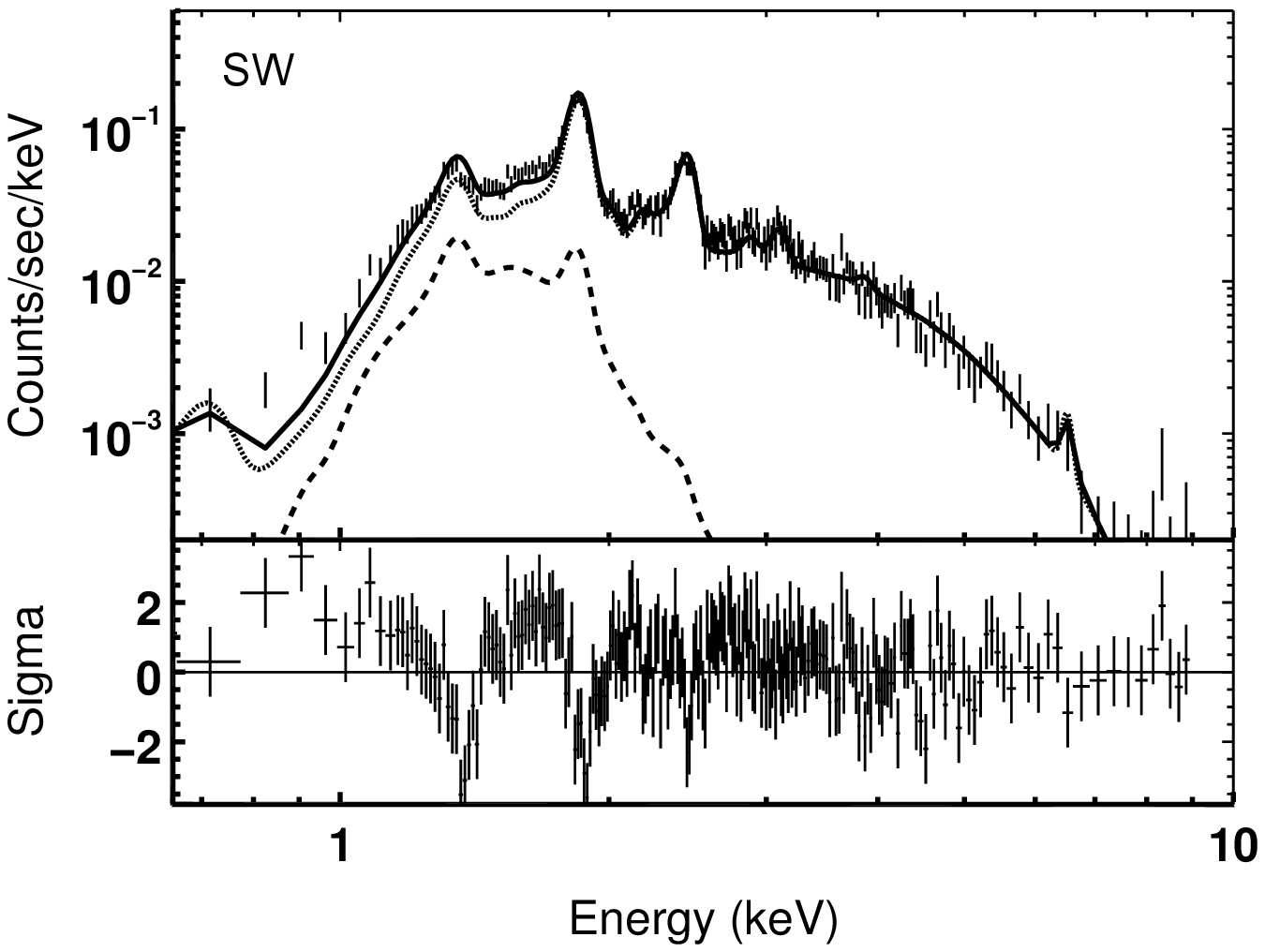} \caption{\label{xrayspec3} Same as Figure \ref{xrayspec1}, but with a two-temperature VNEI model (solid line), composed of a hotter component with a temperature of 1.6 (2.1) keV (dotted line) and a cooler component with a temperature of $\sim$ 0.2 (0.2) keV (dashed line) for the SE (SW) shell.}
\end{figure*}

\begin{figure}
\epsscale{1.1} \plotone{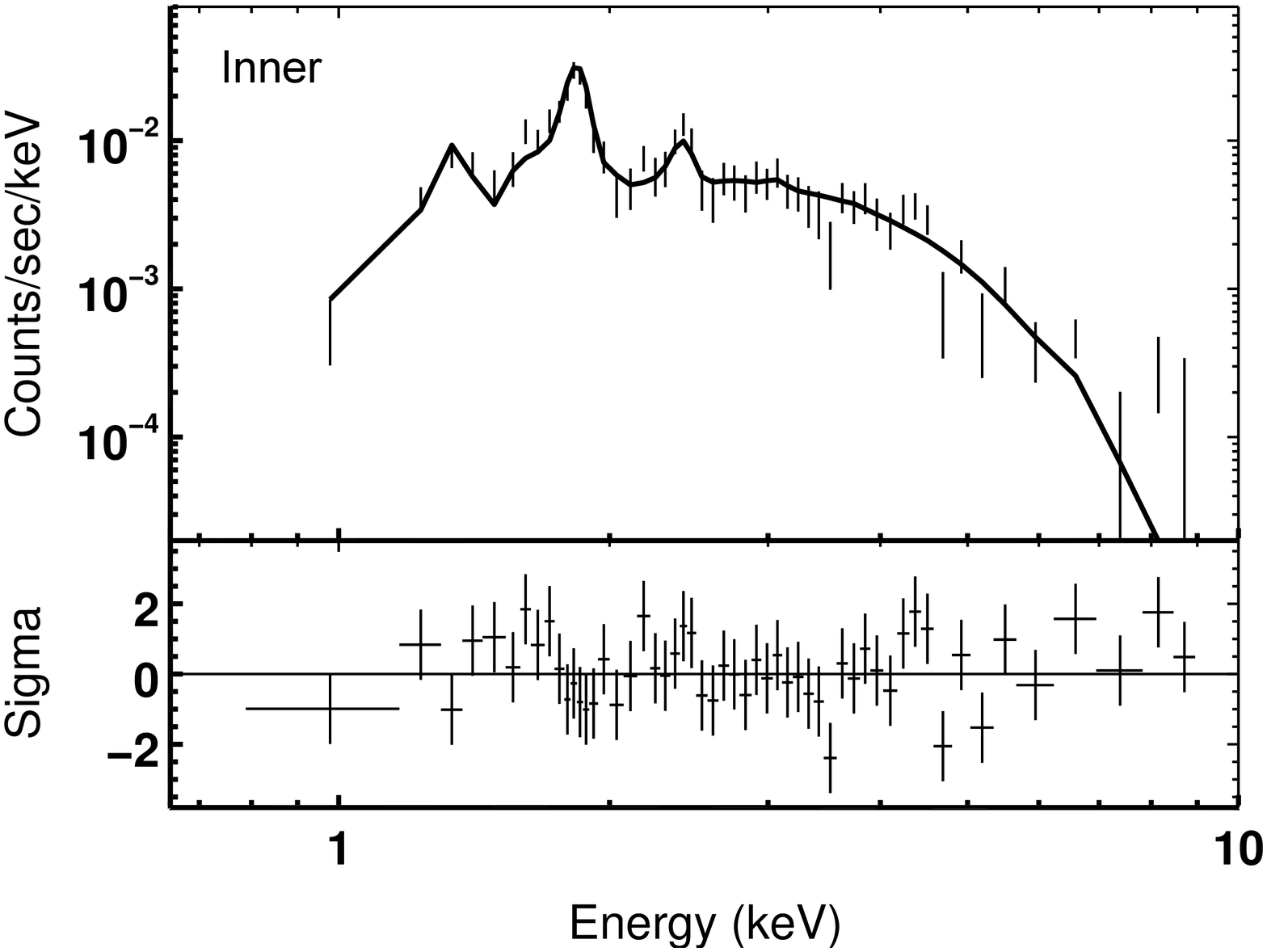} \caption{\label{xrayspecin} \textit{Chandra} X-ray spectra and best-fit VPSHOCK model for the inner SNR region shown in Figure \ref{3color2}. The best-fit thermal temperature is $\sim$2.5 keV and the observed emission lines are from Si, S, Mg, Ar, and Fe. Other parameters are listed in Table \ref{xrayfitstab}.}
\end{figure}

The X-ray spectra were fit with two different models, a non-equilibrium ionization collisional plasma model (VNEI) and a plane-parallel shock plasma model (VPSHOCK), that also include elemental abundances as free parameters. The models are shown in 	Figures \ref{xrayspec1}, \ref{xrayspec2}, \ref{xrayspec3}, and \ref{xrayspecin}. A single temperature VNEI thermal component did not provide a good fit to the SE and SW regions, while these areas were well-fit by the single-temperature VPSHOCK model (see Figures \ref{xrayspec1} and \ref{xrayspecin}). The best-fit parameters for all models are listed in Table \ref{xrayfitstab}. The single-temperature VPSHOCK model implies a temperature of 1.48 keV and 1.93 keV in the SE and SW, respectively, and 2.47 keV in the inner SNR region. The ionization timescales are on the order of $\rm 10^{11}\:s\:cm^{-3}$ in the shell and $\rm 10^{10}\:s\:cm^{-3}$ in the inner SNR region. We let the hydrogen column density vary and found that it is somewhat lower than for the PWN ($N_{\rm H}=3\times10^{22}\: \rm cm^{-2}$), as suggested by \citet{su09}. 

As will be discussed in the following sections, the IR data imply the presence of an additional, lower temperature plasma component. A somewhat better $\chi^2$ is achieved using a two-temperature VPSHOCK model, shown in Figure \ref{xrayspec2}. For comparison, we also fit the data with a two-temperature VNEI model (Figure \ref{xrayspec3}), but the fit was slightly worse in this case. The temperature of this second component is approximately 0.2 keV for both models, but the normalization and ionization timescale are not well constrained. A similar $\chi^2$ value is achieved for a relatively short ionization timescale of $\rm \sim10^{9}\:s\:cm^{-3}$  and a very long timescale of $\rm \sim10^{13}\:s\:cm^{-3}$. The $F$-test implies that the presence of this lower temperature component is not statistically significant. However, as will be discussed in the following sections, the IR data imply the presence of an additional, lower temperature plasma component. 

The X-ray spectra show clear evidence for Si, S, Ar, Mg, and Fe lines, but in contrast to previous studies \citep{mor07,su09}, we find no definitive evidence for enhanced abundances in either of the thermal components. The main difference in our analysis is a more careful approach in background selection that accounted for the emission from the dust scattering halo. After subtracting this emission from the source spectra, the enhanced Fe abundance was no longer required by the fit. While the best fit residuals in Figures \ref{xrayspec1}, \ref{xrayspec2}, and \ref{xrayspec3} clearly show that the models do not provide a good fit to some of the emission lines, varying the abundances did not significantly improve the fit, and the abundances remained around the solar value. We believe that the residuals are caused by the variation in temperature in the higher energy component with a shorter ionization timescale. These differences in temperature are clearly seen in Figure \ref{3color2}(a), and indeed, fits to small sub-regions in the SE and SW shell show that the temperature ranges from 0.9 keV for some of the smaller clumps, up to 2 keV for more diffuse regions. The abundances were consistent with solar even for these various sub-regions.


\subsection{Infrared Spectroscopy}\label{irspec}

\begin{deluxetable*}{lccccc}
\tablecolumns{6} \tablewidth{0pc} \tablecaption{\label{spitzertab}IRS Line Fits}
\tablehead{
\colhead{Line ID} & \colhead{Line Center} & \colhead{Line Flux} & \colhead{De-reddened Flux} & \colhead{FWHM} & \colhead{FWHM} \\
\colhead{} & \colhead{(\micron)} & \colhead{($10^{-6} erg\:cm^{2-}\:s^{-1}\:sr^{-1}$)} & \colhead{($10^{-6} erg\:cm^{2-}\:s^{-1}\:sr^{-1}$)} & \colhead{(\micron)} & \colhead{($km\:s^{-1}$)}
}
\startdata
\sidehead{LH east}
$[$\ion{O}{4}$]$ (25.8903) & 25.8914 $\pm$ 0.0012 & 13.47 $\pm$ 1.03 & 31.52 $\pm$ 2.41 & 0.047 $\pm$ 0.001 &  544 \\
$[$\ion{Fe}{2}$]$ (25.9883) & 25.9751 $\pm$ 0.0013 & 2.86 $\pm$ 0.25 & 6.66 $\pm$ 0.58 & 0.044 $\pm$ 0.001 &  511 \\
$[$\ion{S}{3}$]$ (33.4810) & 33.4733 $\pm$ 0.0019 & 37.81 $\pm$ 4.51 & 71.84 $\pm$ 8.57 & 0.051 $\pm$ 0.002 &  458 \\
$[$\ion{Si}{2}$]$ (34.8152) & 34.8113 $\pm$ 0.0011 & 37.13 $\pm$ 2.11 & 68.80 $\pm$ 3.97 & 0.056 $\pm$ 0.001 &  482 \\
\sidehead{LH west}
$[$\ion{O}{4}$]$ (25.8903) & 25.9009 $\pm$ 0.0019 & 4.87 $\pm$ 0.58 & 11.39 $\pm$ 1.36 & 0.049 $\pm$ 0.002 &  564 \\
$[$\ion{S}{3}$]$ (33.4810) & 33.4816 $\pm$ 0.0014 & 24.06 $\pm$ 1.94 & 45.71 $\pm$ 3.69 & 0.054 $\pm$ 0.001 &  483 \\


\enddata
\tablecomments{The listed uncertainties are 1$\sigma$ statistical uncertainties from the fit only. Due the highly variable background in the vicinity of Kes 75, the measured line intensities should be treated as upper limits.}
\end{deluxetable*}

The IRS slits were positioned along the brightest regions of the SW and SE parts of the shell (see Figure \ref{slits}). The final background subtracted LH spectra of the shell are shown in Figure \ref{highres}. The spectra of the shell are dominated by continuum emission from dust with very little contribution from line emission. Emission lines that are present in the background-subtracted LH spectra are listed in Table \ref{spitzertab}, and they include [{\ion{O}{4}] (25.89 $\micron$), [\ion{Fe}{2}] (25.99 $\micron$), [\ion{S}{3}] (33.48 $\micron$), and [\ion{Si}{2}] (34.82 $\micron$). The listed uncertainties in the observed line intensities are statistical uncertainties only, and they do not include the uncertainties caused by the spatially variable background in the vicinity of Kes 75. Since the background for the high-resolution spectra was extracted from a single region south of the SNR (see Figure \ref{slits}), it is difficult to say how much line emission, if any, is associated with Kes 75. For this reason, the measured line intensities should be treated as upper limits. All of the lines except \ion{O}{4} appear in the local background. In addition, the lines all have a full width at half-maximum (FWHM) that is similar to the expected resolution of the IRS of 500 $\rm km\:s^{-1}$, and none show evidence for significant broadening that would be expected from a rapidly expanding blast wave. There are no lines detected in the LL modules after background subtraction. However, the spectra have high uncertainties due to the high local background that varies significantly across the LL slits.

Extinction correction was applied to the emission line intensities and the LL spectra used in the dust model fitting. We used the extinction curve of \citet{chi06}, and the relation $N_{\rm H}/A_{K}=1.821\times10^{22}{\rm\ cm^{-2}}$ from \citet{dra89}. We set the value of $N_{\rm H}=3.5\times 10^{22}\: \rm cm^{-2}$, an average derived from the fitting of the \textit{Chandra} data to the SE and SW shell.

\subsection{Modeling of the Dust Emission}\label{modeling}

The morphological similarities between the X-ray and IR emission in the southern part of the shell suggest that the emitting dust is likely collisionally heated by the shocked, X-ray emitting gas. The lack of strong IR line emission also indicates that no cooler and denser regions are present where strong UV and optical emission might radiatively heat the dust. The dust grain model used to fit the IR spectra is the BARE-GR-FG model from \citet{zub04} that includes polycyclic aromatic hydrocarbons (PAHs), silicate, and graphite grains. \citet{zub04} determined the grain size distribution for the model by simultaneously fitting the average interstellar extinction, emission, and abundances. Our fitting was performed using a dust heating code that calculates spectral models for dust grains immersed in an X-ray plasma \citep{dwe86,are10}. The model includes stochastic heating of smaller grains and thermal sputtering of dust grains that reduces the size of the initial grain population. The free parameters of this dust model are the electron density $n_e$, electron temperature $T_e$, and the change in the size of each grain due to sputtering, $\Delta a$, that depends on the age of the shocked gas. We produced a grid of models using a density range $n_e\:\rm(cm^{-3})$=[0.35, 0.7, 1.4, 2.8, 5.7, 11, 23, 45, 90, 180], and $T_e\:\rm(K)$=[1.0,1.8, 3.2, 5.6, 10, 18, 32]$\times 10^{6}$. The sputtering amount $\Delta a$ was varied from 0.0--0.01 $\micron$ in steps of 0.0005 $\micron$.

We used the above models to fit spectra from three different regions along the Kes 75 shell; the brightest SE and SW rims and the fainter diffuse emission inside of the shell. The LL sub-slits used for spectral extractions are shown in Figure \ref{3color2}(b). We initially used a background from outside of the SNR for all three regions, and we found that the spectra from the SE and SW cannot be well fitted by a single set of $n_e$ and $T_e$ parameters. The emission appears to originate from two distinct dust temperatures, a warmer component, and a cooler one that is more spatially extended. The emission from this cooler dust also fills the inside of the SNR (``Inner SNR" region in Figure \ref{3color2}). In order to isolate the warmer dust component that is only present in the rims, we decided to subtract the inner SNR contribution from the SE and SW spectra and then fit the spectra with a single temperature. The final fitting was carried out for regions ``SE" and ``SW" in Figure \ref{3color2}(b) where region labeled ``Inner" was used as the background. This inner region was then fitted separately using a background region from outside of the SNR shell. The spectra are shown in Figures \ref{dust_model} and \ref{dust_model2}. Prior to fitting, emission lines were subtracted and the spectra were extinction corrected using the extinction curve from \citet{chi06} and $N_{\rm H}=3.5\times10^{22}\: \rm cm^{-2}$.
The best-fit models for the SE and SW regions are shown as solid red curves in Figure \ref{dust_model}, while the best-fit model for the inner SNR region is shown in Figure \ref{dust_model2}. The dotted red curve in Figure \ref{dust_model} represents the spectrum of the grain size distribution prior to any sputtering. The best-fit parameters for these models are listed in Table \ref{dustfits}, while the $\chi^2$ contour levels for these parameters are shown in Figure \ref{chiplots}, where the shaded regions represent values of [1.2,2,4,8,16,32]$\times$ min $\chi^2$. 

\begin{figure}
\epsscale{1.0} \plotone{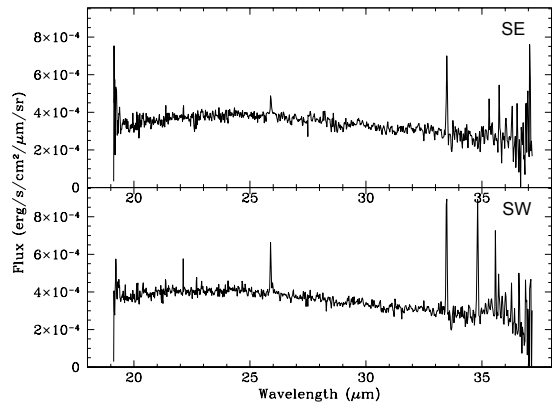} \caption{\label{highres} Background-subtracted high-resolution \textit{Spitzer} IRS LH spectra of the southeastern (top) and southwestern (bottom) portions of the shell in Kes 75. The background was extracted from the southernmost LH slit in Figure \ref{slits}. The observed emission lines show no evidence for broadening and most likely originate from the background. The emission line intensities are listed in Table \ref{spitzertab}.}
\end{figure}


\subsection{Gas and Dust Parameters Derived from IR Emission}

 \begin{figure*}
\epsscale{1.0} \plotone{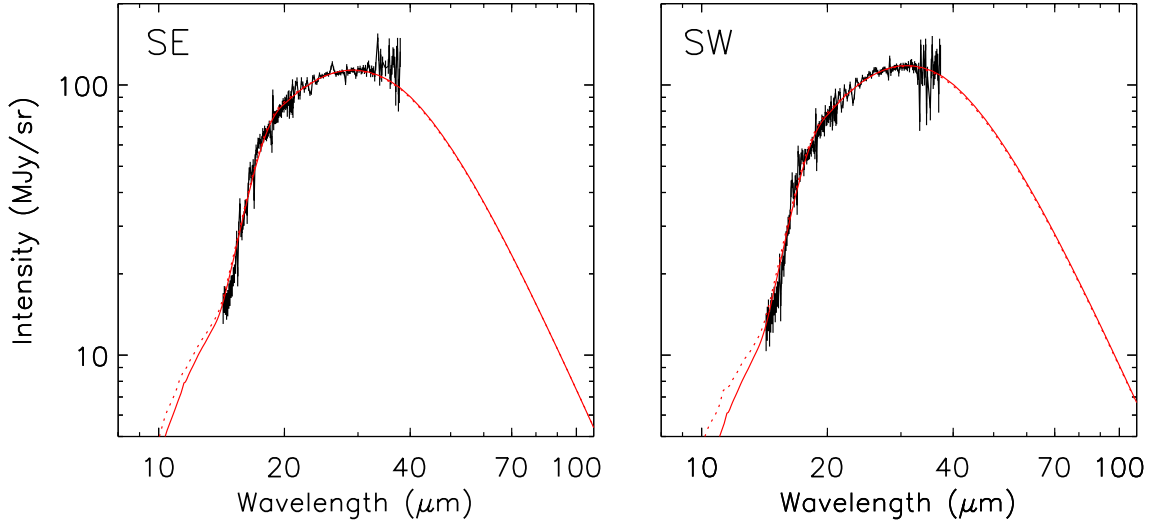} \caption{\label{dust_model} Dust fitting results to the low-resolution IRS spectra of the SE and SW rims of the SNR shell, with the inner SNR region as background. The extraction slits are shown in Figure \ref{3color2}. The spectra have been line subtracted and extinction corrected using the extinction curve from \citet{chi06}. The solid (dotted) red line shows the best-fit model with (without) sputtering. The parameters for the best-fit models are listed in Table \ref{dustfits}. The temperature of the emitting dust in these regions is 140~K.}
\end{figure*}

\begin{figure}
\epsscale{1.0} \plotone{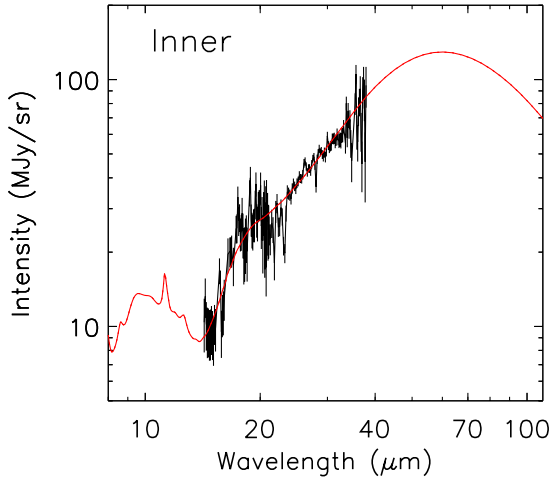} \caption{\label{dust_model2} Dust fitting results to the low-resolution IRS spectra of the inner SNR region, shown in Figure \ref{3color2}(b). The spectrum has been line subtracted and extinction corrected using the extinction curve from \citet{chi06}. The solid red line shows the best-fit model, with parameters listed in Table \ref{dustfits}. The  temperature of the emitting dust in these regions is 55~K.}
\end{figure}

The diffuse IR emission from the inner SNR region seems to originate from a dust population in which the larger grains are characterized by a temperature of 55~K, while the emission from the SNR shell requires the 55~K component in addition to a warm dust component with a temperature of 140~K. We used the inner SNR region as a background for the SE and SW shell spectra, and in that way we have subtracted the contribution from the more diffuse, cooler dust component. The residual IR spectrum from the warm dust in SE and SW rims requires a high plasma temperature and relatively high electron density. The best fit value for the electron density for both regions is $n_e=90\: \rm cm^{-3}$, with an electron temperature of $T_e=(1--2)\times10^{7}\: \rm K$ (0.9--1.7 keV). While the fit clearly requires a relatively high density, the temperature is somewhat less constrained (see Figure \ref{chiplots}).

The best fit values for the sputtering parameter $\Delta a$ are 0.01 (0.002) $\micron$ for the SE (SW) spectrum. These values imply that almost all of the smaller PAH grains have been destroyed and approximately 50\% (25\%) of the mass of silicate and graphite grains. However, since we only have IR spectra longward of 15 $\micron$, this sputtering parameter $\Delta a$ is not well constrained by the model, even though the lack of emission at IRAC wavelengths does imply that the smaller grains have likely been destroyed in these regions. 

While the presence of the cooler plasma component is not conclusive from the fits to the X-ray data, it is implied by the presence of the 55~K dust component. The best fit model to the IR spectrum of the inner SNR region implies a lower electron temperature of $\rm 1.8\times10^6\:K$ (0.16 keV) and a much lower electron density, $n_e=2.8\: \rm cm^{-3}$. The minimum $\chi^2$ for the inner region is actually achieved for a broader density range of 1--15 $\rm cm^{-3}$, but this is still several times lower than what is required for the SE and SW rims. The $\Delta a$ parameter is better constrained for the inner region. The best-fit value of $\Delta a=0$ implies that very little sputtering has occurred in this region (see Figure \ref{chiplots}).

\begin{deluxetable*}{lcccc}
\tablecolumns{5} \tablewidth{0pc} \tablecaption{\label{dustfits}Dust Model Fits to IRS Spectra}
\tablehead{
 \colhead{Parameter} & \colhead{$n_e \rm(cm^{-3})$} & \colhead{$T_e\rm \:(K)$} & \colhead{$\rm \chi^2$} & \colhead{$\rm Mass\:(M_{\odot})$} 
}
\startdata
\cutinhead{Parameters from IR Fits}
SE shell & 90 & $1.8\times10^7$ (1.55 keV) & 2.3 & 6$\times10^{-3}$ (3$\times10^{-3}$) \\
SW shell & 90 & $1.0\times10^7$ (0.86 keV) & 3.2 & 7$\times10^{-3}$ (5$\times10^{-3}$) \\
Inner SNR & 2.8 & $1.8\times10^6$ (0.15 keV) &  2.6 & 0.9 (SE) \\
&  &  &  & 1.1 (SW) \\
\cutinhead{Parameters from X-Ray Fits (VPSHOCK)}
SE hot & 5.6 & $1.7\times10^7$ (1.46 keV) & & 1.7 \\
SW hot & 3.1 & $1.9\times10^7$ (1.60 keV) & & 1.7 \\
SE warm & $<$138  &  $2.3\times10^6$ (0.20 keV) & & $< $43 \\
SW warm & $<$200 & $1.6\times10^6$ (0.14 keV) & & $<$110 \\
\cutinhead{Parameters from X-Ray Fits (VNEI)}
SE hot & 5.0 & $1.9\times10^7$ (1.60 keV) & & 1.5 \\
SW hot & 2.3 & $2.2\times10^7$ (2.08 keV) & & 1.2 \\
SE warm & 23  &  $2.8\times10^6$ (0.24 keV) & & 7.0 \\
SW warm & $<$40 & $2.3\times10^6$ (0.20 keV) & & $<$20 \\
\cutinhead{Dust-to-Gas Ratios for the Hot Component}
SE hot & & & & 4$\times10^{-3}$ ($2\times10^{-3}$) \\
SW hot & & & & 4$\times10^{-3}$ ($3\times10^{-3}$) \\
\enddata
\tablecomments{The masses listed for the IR fits are dust masses, while the masses listed for the X-ray fits are estimated gas masses. The dust temperatures are 140 K (warm component) and 55 K (cool component). The densities from the X-ray fits were calculated from the normalization of the VPSHOCK model and assuming a filling factor f $=$ 1. Dust masses and dust-to-gas mass ratios outside (inside) parentheses represent the best-fit values without (with) sputtering.}
\end{deluxetable*}

\section{DISCUSSION}

\subsection{Comparison of Derived IR and X-Ray Parameters}

Interestingly, based on the best-fit electron temperature to the SE and SW IR spectra, the dust in these regions appears to be associated with the higher temperature (2$\times10^7$ K) X-ray emitting plasma, rather than the 0.2 keV (2$\times10^6$ K) component as suggested by \citet{mor07}. The association of the IR emission with this harder X-ray emission is also implied by the spatial correlation of the IR and X-ray emission, as seen in Figure \ref{3color2}(b). The IR emission from the inner SNR region may be associated with the lower temperature plasma component, since the IR and X-ray fits provide independent estimates of the $n_e$ and $T_e$ that are consistent with each other.

The electron density for the two thermal components in the X-ray fits was determined from the normalization of the VPSHOCK model, $10^{-14}n_e n_{\rm H} V / 4\pi d^2 \: \rm cm^{-5}$, where V is the volume of the emitting region and $d$ is the distance to the SNR, set to the value of 10.6 kpc. Similar to \citet{mor07}, we assumed a slab-like volume for the X-ray emitting regions. We took the area of the SE and SW regions shown in Figure \ref{3color2}(a) and assumed that the depth along the line of sight is equal to the long axis of each region. For an assumed distance of 10.6 kpc, the assumed volumes are $\rm 15(26)\: pc^3$ for the SE(SW) region. We also assume that $n_e=1.18 n_{\rm H}$, as for a fully ionized gas. The estimated electron densities for the X-ray emitting components are $5.6f^{-1/2}d^{-1/2}_{10.6}\rm \:cm^{-3}$ and $3.1f^{-1/2} d^{-1/2}_{10.6}  \:cm^{-3}$ for the $\sim$ 1.5 keV component for SE and SW rims, respectively (Table \ref{dustfits}), where $f$ is the filling factor that accounts for the fact that the emitting gas may occupy a smaller fraction of the chosen volume. The ionization timescale implies a similar density, assuming that the shock age is equal to the estimated SNR age of 880 yr. Since the normalization of the cooler plasma component is not constrained, we can only derive upper limits on the density and gas mass in the $\sim $0.2 keV component. The upper limits on the density and mass calculated from the normalization of the VPSHOCK model (best-fit value plus 1.6$\sigma$) for the same spatial regions (SE/SW regions in Figure \ref{3color2}(a)) are $140d^{-1/2}_{10.6}\rm \:cm^{-3}$ and $ 200d^{-1/2}_{10.6}\: \rm cm^{-3}$, and $43d^{5/2}_{10.6} M_{\odot}$ and $110d^{5/2}_{10.6} M_{\odot}$ for the SE and SW, respectively. These values are listed in Table \ref{dustfits}.

The estimate of $n_e$ from the X-ray fits is significantly lower than the high density derived from the fits to the IR spectrum. If the warm dust component is associated with the 1.5 keV plasma, as suggested by implied temperatures, it may be possible that the X-ray emitting gas in the shell is clumpy and concentrated in a smaller volume. If we assume that the density of the clumps is $\rm 90 \: cm^{-3}$, as derived from the IR data, we estimate a low filling factor that is on the order of a few times $10^{-3}$. As can be seen from Figure \ref{3color2}(a), the shell does indeed appear to be clumpy, so assuming that higher density clumps are present in the shell may not be unreasonable. However, the large discrepancy between the X-ray and IR derived densities remains a problem.

\subsection{Dust Mass and Dust-to-Gas Mass Ratio}\label{dusttogas}

We estimated the total gas masses in the hot component of the shell to be $1.7d^{5/2}_{10.6}\: M_{\odot}$ in each of the rims. The dust masses were calculated using a grain size distribution from the best fit models and the total MIPS 24 $\micron$ fluxes from the SE and SW rims of the shell. The fluxes used for the total dust estimates are listed in Table \ref{irfluxtab}. For the warmer dust component we used residual fluxes after subtracting the contribution from the inner SNR emission. These values are equal to 2.2 Jy (SE) and 3.4 Jy (SW). We then subtracted these values from the total flux in the shell to estimate the contribution of the cooler dust component and find these values to be 3.3 Jy (SE) and 3.6 Jy (SW). Dust mass estimates are listed in Table \ref{dustfits}. We find a total dust mass of $6\times 10^{-3}d^{2}_{10.6} \: M_{\odot}$ (SE) and $7\times 10^{-3}d^{2}_{10.6} \: M_{\odot}$ (SW) for the 140~K dust component, and $0.9d^{5/2}_{10.6}\: M_{\odot}$ (SE) and $1.1d^{5/2}_{10.6} \: M_{\odot}$ (SW) for the 55~K component. Since the sputtering amount for the hot dust component is not constrained by our model, the calculated dust masses assume no sputtering and should be treated as upper limits. The best-fit dust mass values that include sputtering are listed in Table \ref{dustfits}. The estimate of the mass of the cooler dust component is highly uncertain since we assume that the diffuse IR emission all comes from dust with a temperature of 55 K, even though this may be true only for the region where we have an IR spectrum. A lower (higher) dust temperature would reduce (increase) the mass associated with this component.

We only estimate the dust-to-gas mass ratio for the 140~K dust since the evidence suggests that it is associated with the hot X-ray plasma. The estimates of the dust-to-gas mass ratios are listed in Table \ref{dustfits}. Assuming that the warm dust component is associated with the hotter X-ray plasma, we calculate a dust-to-gas mass ratio of $4\times 10^{-3}d^{-1/2}_{10.6}$ for the SE and SW rims. These values are consistent with the dust-to-gas mass ratio of $0.6^{+0.4}_{-0.2}$\% that is typically assumed for the Galaxy \citep{sod94}, especially considering the uncertainties involved in determining the gas and dust masses and the distance to Kes 75. If dust sputtering has occurred in the shell, the dust-to-gas mass ratio would be a few times lower. Since the sputtering amount is not constrained by our spectrum, we cannot conclude whether any significant dust destruction has occurred in the shell. 

\subsection{Origin of the IR and X-Ray Emission}\label{origin}

\begin{figure}
\epsscale{1.0} \plotone{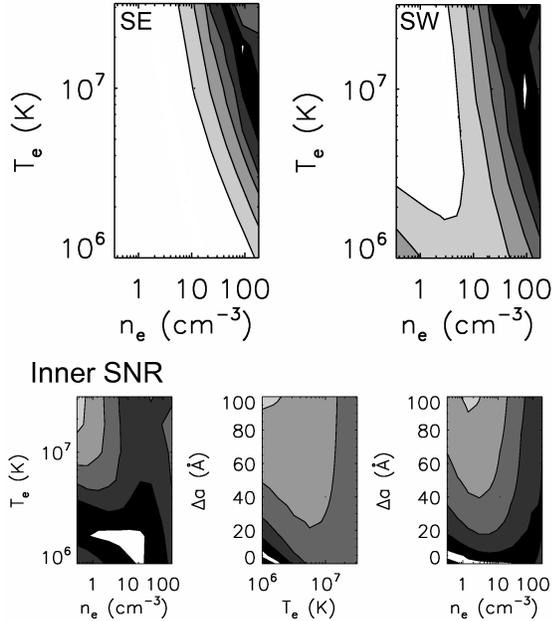} \caption{\label{chiplots} Contour levels representing [1.2,2,4,8,16,32]$\times$ min $\chi^2$ for dust model fits for SE, SW, and inner SNR regions. The sputtering parameter $\Delta a$ could not be constrained for the SE and SW shell, so contour plots for $\Delta a$ are not shown for these regions. The best fit parameters for $T_e$ and $n_e$ are listed in Table \ref{dustfits}.}
\end{figure}

\citet{mor07} attributed the low-temperature X-ray thermal component to the forward-shocked circumstellar medium (CSM)/ISM, and the high-temperature component to the reverse-shocked ejecta. Our modeling of the IR spectrum implies that the hotter temperature plasma is required to collisionally heat the dust in the shell to the observed temperature of $\sim$ 140~K. This association is also implied by the similarities in the morphology between the IR emission and the hotter X-ray gas. If the hot X-ray component does indeed originate from the reverse-shocked ejecta, then the warm dust in the shell may be coming from ejecta dust produced in the SN explosion. The total mass of this dust is approximately $1.3\times 10^{-2}d^{2}_{10.6}\: M_{\odot}$, not unreasonable for such an interpretation. However, the shape of the IR spectrum shows that the composition of the dust is consistent with that of the ISM, and it does not show any features typical of ejecta dust in other SNRs observed with \textit{Spitzer}, such as Cas A and G54.1+0.3 \citep{rho08,tem10}. Since we see no clear evidence for enhanced abundances in the X-ray spectra, it is possible that the range of different plasma temperatures and densities are due to the interaction of the SNR blast wave with the expansion into a non-uniform ambient medium or clumpy CSM. In this case, the observed dust emission originates from the swept-up CSM/ISM dust. The emission from the inner SNR in this case would be projected emission from the forward-shocked material along the line of sight. 

 The modeling of the IR spectra from the inner region of the SNR implies that a gas component with a temperature of less than 0.2 keV needs to be present to heat the dust to a temperature of 55~K. This is consistent with the temperature of the second thermal component allowed by the X-ray fits in Table \ref{xrayfitstab}, thus favoring the two-component thermal model for the X-ray spectrum. The dust temperature is too warm to be radiatively heated by the PWN, so collisional heating of the dust by the forward-shocked ambient gas seems to be the most plausible explanation. For a typical Galactic dust-to-gas mass ratio, the rough estimate of the cooler dust mass of $\sim 2d^{2}_{10.6} \: M_{\odot}$ implies that the associated gas mass would need to be very high, on the order of $\sim 280 d^{2}_{10.6}\: M_{\odot}$. However, if we assume a lower value for the distance, or a higher average dust temperature for the diffuse IR emission, this would significantly reduce the mass estimate of this gas component. The X-ray data do not exclude the possibility of a more massive cooler gas component, because its temperature may be too cool to be detected in X-rays due to the high absorbing column density toward Kes 75. There is also a possibility that the more diffuse IR emission from the cooler dust is not associated with the remnant even though it appears to be morphologically contained within the SNR shell.

\section{Conclusion}
 
 We presented the analysis of \textit{Chandra} X-ray and \textit{Spitzer} IR spectroscopy of the Kes 75 SNR shell. We find that the X-ray data are well fitted by a $\sim$ 1.5 keV VPSHOCK model, but we find no clear evidence for enhanced abundances that would be expected from metal-enriched ejecta. In addition, the modeling of IR spectra and the observed morphology suggest that at least $1.3\times10^{-2}\:M_{\odot}$ of warm dust (140~K) is associated with the thermal plasma seen with \textit{Chandra}. Since the amount of sputtering in the shell is not constrained by our model, we cannot conclude if significant dust destruction is occurring in the shell. If the hot plasma is associated with the reverse-shocked ejecta, the dust is likely SN-formed dust. However, the emission from the gas and dust may also be associated with the interaction of the forward shock with a clumpy CSM. There is a discrepancy between the densities derived from the X-ray and IR data. While the X-ray data imply a low density on the order of $\rm 1--10\:cm^{-3}$, the IR model points to a much higher density of $\rm 90\:cm^{-3}$. A low filling factor for the emitting gas may help reconcile these values, however the required factor appears unusually low. 
 
 The IR data also show a more diffuse cooler dust component present in the SNR. The modeling of the IR spectra imply that a cooler plasma component with $n_e=2.8\: \rm cm^{-3}$ and $T_e=1.8\times 10^6\:\rm K$ (0.15 keV) is required to heat this dust to the observed temperature of 55~K. The dust mass in this component is $\sim2d^{2}_{10.6} \:M_{\odot}$ and the associated gas mass is $\sim280d^{2}_{10.6} \:M_{\odot}$, assuming a Galactic dust-to-gas mass ratio. The mass estimates of the cooler dust and gas components are lower if we assume a smaller distance to Kes 75, or a higher average dust temperature for the diffuse IR emission in the shell. The X-ray data do not discount this possibility, especially if this component is too cool to be detected with the high hydrogen column density of $N_{\rm H}=(3--4)\times10^{22}\: \rm cm^{-2}$. While we cannot make any definite conclusions about the nature of the X-ray emitting gas, this study highlights the importance of IR data in deriving gas and dust properties. Future observations of Kes 75 with \textit{Herschel} and \textit{James Webb Space Telescope}, especially determining the spatial extent and temperature of the cooler dust component and obtaining IR spectra below 15 $\micron$ that will constrain the sputtering amount, would help resolve some of the discrepancies and shed light on the origin of the IR and X-ray emission in the Kes 75 shell.
 
\acknowledgments
This work is partly based on observations made with the \textit{Spitzer Space Telescope}, which is operated by the Jet Propulsion Laboratory, California Institute of Technology under a contract with NASA. Support for this work was provided by NASA through an award issued by JPL/Caltech (RSA1343487). The research has made use of software provided by the CXC in the application packages CIAO, ChIPS, and Sherpa. P.O.S. acknowledges partial support from NASA Contract NAS8-03060.
We acknowledge Stephen Reynolds and Kazik Borkwoski at North Carolina State University for the useful discussion and suggestions, and George Sonneborn at NASA GSFC for helpful comments.

\bibliographystyle{plain}

\end{document}